# Prerequisite for superconductivity: appropriate spin-charge correlations

Tian De Cao[*]

*Department of physics, Nanjing University of Information Science & Technology, Nanjing 210044, China*

**Abstract**

This work investigates the relation between superconductivity and correlations. A simple calculation shows that the appropriate spin-charge correlation is the key role to any superconductivity, and this calculation is consistent with the analyses of unusual properties of superconductors. (Note: the Tc of this model is not given clearly in this work, but we have advanced this mechanism to a t-χ model which includes various superconductivities and magnetisms (please see arXiv:0707.3660 and following works).)

**PACS:** 74.25.Jb; 74.72.-h; 74.62.Yb.

**Keywords:** D. Spin correlation; D. Charge correlation; D. Superconductivity. arXiv:0707.366

## 1. Introduction

Since mother compounds of the cuprate superconductors are typical Mott insulators with antiferromagnetic order, the relation between magnetism and superconductivity has been noted in many works. Schrieffer et al. give their spin bag model in 1988[1]. Other theories concerning spins include spin polaronic theories [2,3,4] and antiferromagnetic spin fluctuation based mechanisms [5,6,7,8]. Although the coexistence between magnetism and superconductivity have also been investigated by other works [9,10], how spins contribute to superconductivity has been an open and important problem. In this work, by introducing spin operator and charge operator into the Hubbard model to calculate some correlation functions, we can find the roles of different correlations. It is shown that the appropriate spin-charge correlation is the key factor of the superconductivity; this explains why the materials appearing the larger magnetic susceptibility $|\chi_m|$ (but a lesser electronic susceptibility) usually show superconductivity, and why the high-temperature superconductivities occur at the region where the short-range antiferromagnetic order [11] exists in the cuprate superconductors.

## 2. Calculation

To consider the roles of strong correlations, we discuss the Hamiltonian

$$\hat{H} = \sum_{l,l',\sigma} T_{ll'} c^+_{l\sigma} c_{l'\sigma} + U\sum_{l} c^+_{l\uparrow} c_{l\uparrow} c^+_{l\downarrow} c_{l\downarrow} + \sum_{l,l',\sigma,\sigma'} V_{ll'} c^+_{l\sigma} c_{l\sigma} c^+_{l'\sigma'} c_{l'\sigma'}, \qquad (1)$$

where $c^+_{l\sigma}$ creates as electron at site $l (\equiv \vec{R}_l)$ of spin $\sigma$ in the model, $t_{ll'}$ is the intersite hopping matrix element, $U$ the effective on-site Coulomb interaction, and $V_{ll'}$ the long-range interaction, particularly, the nearest-neighbor interaction. We will find that $U$ and $V_{ll'}$ dominate short- and long-range correlations, respectively.

To find the effects of spins and charges, we reduce the model in the form

$$\hat{H} = \sum_{l,l',\sigma} T_{ll'} c^+_{l\sigma} c_{l'\sigma} + U\sum_{l} (\hat{\rho}_l^2 - \hat{s}_l^2) + 4\sum_{l,l'} \hat{\rho}_l V_{ll'} \hat{\rho}_{l'}, \qquad (2)$$

---

[*]Corresponding author.
[*]E-mail address: tdcao@nuist.edu.cn (T. D. Cao).
[*]Tel: 0086-25-58356487

using

$$\hat{\rho}_l = \frac{1}{2}\sum_\sigma c_{l\sigma}^+ c_{l\sigma}, \quad \hat{s}_l = \frac{1}{2}\sum_\sigma \sigma c_{l\sigma}^+ c_{l\sigma}. \tag{3}$$

The spin operator is $\hat{s}_l \equiv \hat{s}_{lz}$, $\sigma = \pm 1$ represent spin up and spin down, and the relation between charge operator $\hat{\rho}_l$ and number operator $\hat{n}_l$ is $\hat{\rho}_l = \hat{n}_l/2$. It is found that the on-site interactions include the part of spin-spin interactions besides the charge-charge interactions. Hence spin character, and then some magnetism, is included in the Hubbard model. This can be understood, because the interaction $c_{l\uparrow}^+ c_{l\uparrow} c_{l\downarrow}^+ c_{l\downarrow}$ means that the spin of an electron is in the opposite direction of another one during the scattering process, hence the spin of electrons may be reversed in the scattering process.

To discuss the possible antiferromagnetism, we divide the lattice into A-sublattice and B-sublattice, and write the model (2) as

$$\hat{H} = \sum_{l_A,l_B,\sigma} T_{l_A l_B}(c_{l_A\sigma}^+ c_{l_B\sigma} + c_{l_B\sigma}^+ c_{l_A\sigma}) + U\sum_{l_A}(\hat{\rho}_{l_A}^2 - \hat{s}_{l_A}^2) + U\sum_{l_B}(\hat{\rho}_{l_B}^2 - \hat{s}_{l_B}^2) + 4\sum_{l_A,l_B}\hat{\rho}_{l_A} V_{l_A l_B} \hat{\rho}_{l_B}. \tag{4}$$

Following Mahan [12], these Green's functions are defined in

$$G_{AA}(l_A l_A'\sigma, \tau-\tau') = -<T_\tau c_{l_A\sigma}(\tau) c_{l_A'\sigma}^+(\tau')>, \quad F_{AA}^+(l_A l_A'\sigma, \tau-\tau') = <T_\tau c_{l_A\sigma}^+(\tau) c_{l_A\bar\sigma}^+(\tau')>.$$

From the definition of the $\tau$-ordered product, we derive the first derivative of the equations for the Green's functions, such as

$$\frac{\partial}{\partial \tau} G_{AA}(l_A l_A'\sigma, \tau-\tau') = -\delta(\tau-\tau')\delta_{l_A,l_A'} + \sum_{l_B} T_{l_A l_B} <T_\tau c_{l_B\sigma}(\tau) c_{l_A'\sigma}^+(\tau')>$$

$$+ \frac{1}{2}U[<T_\tau \hat{\rho}_{l_A} c_{l_A\sigma} c_{l_A'\sigma}^+(\tau')> + <T_\tau c_{l_A\sigma} \hat{\rho}_{l_A} c_{l_A'\sigma}^+(\tau')> - \sigma<T_\tau \hat{S}_{l_A} c_{l_A\sigma} c_{l_A'\sigma}^+(\tau')> - \sigma<T_\tau c_{l_A\sigma} \hat{S}_{l_A} c_{l_A'\sigma}^+(\tau')>]$$

$$+ 2\sum_{l_B} V_{l_A l_B} <T_\tau c_{l_A\sigma} \hat{\rho}_{l_B} c_{l_A'\sigma}^+(\tau')>.$$

Therefore, to calculate these Green's functions, we must derive more equations of other correlation functions, such as $<T_\tau \hat{\rho}_l c_{l\sigma} c_{l'\sigma}^+(\tau')>$. Moreover, because

$$\frac{\partial}{\partial \tau}[\hat{\rho}_{l_A} c_{l_A\sigma}] = -\sum_{l_B} T_{l_A l_B} \hat{\rho}_{l_A} c_{l_B\sigma} - \frac{1}{2}U[\hat{\rho}_{l_A}\hat{\rho}_{l_A}c_{l_A\sigma} + \hat{\rho}_{l_A} c_{l_A\sigma} \hat{\rho}_{l_A} - \sigma\hat{\rho}_{l_A}\hat{S}_{l_A} c_{l_A\sigma} - \sigma\hat{\rho}_{l_A} c_{l_A\sigma} \hat{S}_{l_A}]$$

$$- 2\sum_{l_B} \hat{\rho}_{l_A} c_{l_A\sigma} V_{l_A l_B} \hat{\rho}_{l_B} - \frac{1}{2}\sum_{l_B\sigma'} T_{l_A l_B} [c_{l_A\sigma'}^+ c_{l_B\sigma'} - c_{l_B\sigma'}^+ c_{l_A\sigma'}] c_{l_A\sigma},$$

to close the relations between these correlated functions, we approximate $\hat{\rho}_l \hat{\rho}_l c_{l\sigma} = <\hat{\rho}_l \hat{\rho}_l> c_{l\sigma}$, and so forth. In the right side of the equation (5), $l_B \neq l_A'$; if we consider the weak antiferromagnetic states, $S_{l_A} = -S_{l_B} \to 0$, we can approximate $G_{BA}(ll'\sigma, \tau-\tau') \to G_{AA}(ll'\sigma, \tau-\tau')$ for $l \neq l'$ in the right side of equation (5). After exact calculations, we obtain

$$[-i\omega_n + \varepsilon_k + \frac{P_A - \sigma S_A + v_l \varepsilon_k}{i\omega_n}] G_A(k\sigma, i\omega_n) = -1 - \frac{u_{A\sigma}}{i\omega_n} + \frac{1}{i\omega_n} V_{A\Delta}(k\sigma) F^+_{AA}(\bar{k}\bar{\sigma}, i\omega_n), \quad (6)$$

$$[-i\omega_n - \varepsilon_k + \frac{1}{i\omega_n}(P_A - \sigma S_A + v_l \varepsilon_k + g_{k\sigma})] \cdot F^+_{AA}(k\sigma, i\omega_n) = -\frac{1}{i\omega_n} V^+_{A\Delta}(k\sigma) G_{AA}(\bar{k}\bar{\sigma}, i\omega_n), \quad (7)$$

where

$$\varepsilon_k = \sum_{l''} T_{ll''} e^{ik\cdot(l''-l)}, \quad V_k = \sum_{l''} V_{ll''} e^{ik\cdot(l''-l)},$$

and we introduced

$$P^{AB}_{\rho\rho}(l_A - l_B) \equiv P^{AB}_{\rho\rho}(\vec{R}_{l_A} - \vec{R}_{l_B}) = <\hat{\rho}_{l_A} \hat{\rho}_{l_B}>,$$

$$P^{AB}_{\rho S}(l_A - l_B) \equiv P^{AB}_{\rho S}(\vec{R}_{l_A} - \vec{R}_{l_B}) = <\hat{\rho}_{l_A} \hat{S}_{l_B}>,$$

$$P_A = U P^{AA}_{\rho\rho}(0) U + U P^{AA}_{SS}(0) U + 4 \sum_{l_B',l_B} V_{l_A l_B} P^{BB}_{\rho\rho}(l_B' - l_B) V_{l_B l_A} + 4 \sum_{l_B} V_{l_A l_B} P^{BA}_{\rho\rho}(l_B - l_A) U,$$

$$S_A = 2 U P^{AA}_{\rho S}(0) U + 4 \sum_{l_B} V_{l_A l_B} P^{BA}_{\rho S}(l_B - l_A) U,$$

$$v_l = 2 \sum_{l_B} V_{l_A l_B} \rho_{l_B}, \quad u_{A\sigma} = [U(\rho_{l_A} - \sigma S_{l_A}) + 2 \sum_{l_B'} V_{l_B' l_A} \rho_{l_B'}],$$

$$g_{k\sigma} = \sum_{k'} [U \varepsilon_k + V_k \varepsilon_k - U \varepsilon_{k'} - V_k \varepsilon_{k'}] G_{AA}(k'\sigma, 0),$$

$$V_{A\Delta}(k\sigma) = -\sum_{k'} [U \varepsilon_{k'} + V_k \varepsilon_{k'} - U \varepsilon_k - V_k \varepsilon_k] F_{AA}(k'\sigma, 0).$$

In these expressions, we denote $k \equiv \vec{k}$, $s_l \equiv <\hat{s}_l>$, and $\rho_l \equiv <\hat{\rho}_l>$. $S_l$ and $\rho_l$ represent spin and charge at each site respectively. In our calculation we have noted $E_{\bar{k}} = E_k$, $V_{\bar{k}} = V_k$, and $P_{\rho s} = P_{s\rho}$. The cell number $N$ has been taken as $N = 1$, otherwise we should take $\sum_k \rightarrow \frac{1}{N} \sum_k$ in calculations. In addition, to consider the chemical potential of electron systems, we should take $\varepsilon_k \rightarrow \varepsilon_k - \mu$ in our discussion. If we take the transition $A \rightarrow B$, we will obtain the equations of functions $G_{BB}$ and $F^+_{BB}$. It is found that $G_{BB}(k\sigma, i\omega_n) = G_{AA}(k\bar{\sigma}, i\omega_n)$, and $F^+_{BB}(k\sigma, i\omega_n) = F^+_{AA}(k\bar{\sigma}, i\omega_n)$ for $S_{l_A} = -S_{l_B}$. The effects of both charge-charge correlations and spin-spin correlations are described by $P_A$, the effects of spin-charge correlations by $S_B$. In contrast with the charge-charge correlations and the spin-spin correlations, the expressions (6) and (7) show that the effects of spin-charge correlations depend on the direction of spins. The effects of zero-range correlations are determined by $U$ the on-site interaction, while the effects of long-range correlations are affected by $V_{ll'}$ the long-range interaction.

Consider the case $V_{A\Delta}(k\sigma) \rightarrow 0$ for $T \rightarrow T^{pair}$, we get

$$G_{AA}(k\sigma,i\omega_n) = \sum_{\nu=\pm} \nu \frac{C_A^{(\nu)}(k\sigma)}{i\omega_n - E_A^{(\nu)}(k\sigma)}, \tag{8}$$

where

$$C_A^{(\nu)}(k\sigma) = \frac{E_A^{(\nu)}(k\sigma) + u_{A\sigma}}{E_A^{(+)}(k\sigma) - E_A^{(-)}(k\sigma)}, \quad E_A^{(\pm)}(k\sigma) = \frac{1}{2}\{\varepsilon_k \pm \sqrt{\varepsilon_k^2 + 4[v_l\varepsilon_k + P_A - \sigma S_A]}\}.$$

Moreover, we obtain

$$F_{AA}^+(k\sigma,i\omega_n) = V_{A\Delta}^+(k\sigma) \sum_{i=1}^{4} \frac{B_A^{(i)}(k\sigma)}{i\omega_n - E_{k\sigma}^{(i)}},$$

where

$$B_A^{(i)}(k\sigma) = \frac{E_{k\sigma}^{(i)} + u_{A\bar{\sigma}}}{\prod_{j=1(\neq i)}^{4}[E_{k\sigma}^{(i)} - E_{k\sigma}^{(j)}]},$$

$$E_{k\sigma}^{(1)} = \frac{1}{2}\{-\varepsilon_k + \sqrt{\varepsilon_k^2 + 4(P_A - \sigma S_A + v_l\varepsilon_k + g_{k\sigma})}\}, \quad E_{k\sigma}^{(2)} = \frac{1}{2}\{-\varepsilon_k - \sqrt{\varepsilon_k^2 + 4(P_A - \sigma S_A + v_l\varepsilon_k + g_{k\sigma})}\},$$

$$E_{k\sigma}^{(3)} = \frac{1}{2}\{\varepsilon_k + \sqrt{\varepsilon_k^2 + 4(P_A + \sigma S_A + v_l\varepsilon_k)}\} = E_{k\bar{\sigma}}^{(+)}, \quad E_{k\sigma}^{(4)} = \frac{1}{2}\{\varepsilon_k - \sqrt{\varepsilon_k^2 + 4(P_A + \sigma S_A + v_l\varepsilon_k)}\} = E_{k\bar{\sigma}}^{(-)}. \tag{9}$$

Take the transition $\varepsilon_k \to \varepsilon_k - \mu$, and $E_{k\sigma}^{(i)} \to \xi_{k\sigma}^{(i)}$, we get

$$F_{AA}^+(k\sigma,0) = V_{A\Delta}^+(k\sigma) \sum_{i=1}^{4} B_A^{(i)}(k\sigma) n_F(\xi_{k\sigma}^{(i)}),$$

or

$$V_{A\Delta}^+(k\sigma) = -\sum_{k'}[U\xi_{k'} + V_k\xi_{k'} - U\xi_k - V_{k'}\xi_k] \cdot V_{A\Delta}^+(k'\sigma) \sum_{i=1}^{4} B_A^{(i)}(k'\sigma) n_F(\xi_{k'\sigma}^{(i)}). \tag{10}$$

If we take $V_{ll'} = V_1$ and $t_{ll'} = t_1$ for the nearest-neighbor approximation, we get

$$V_{A\Delta}^+(k\sigma) = -(U + 2\mu\frac{V_1}{t_1})\sum_{k'}[\varepsilon_{k'} - \varepsilon_k] V_{A\Delta}^+(k'\sigma) \cdot \sum_{i=1}^{4} B_A^{(i)}(k'\sigma) n_F(\xi_{k'\sigma}^{(i)}), \tag{11}$$

Where $\mu$ is the chemical potential of the electron systems. We can write $I_{t\Delta}^+(k\sigma)$ of the identity (11) as

$$V_{A\Delta}^+(k\sigma) = V_{A\Delta}^+(0\sigma,T) + \gamma_\sigma(T)\varepsilon_k, \tag{12}$$

When $V_{A\Delta}^+(k\sigma) \neq 0$, the substitution of (12) into (11) leads to

$$\sum_{k',k''}(\varepsilon_{k'}\varepsilon_{k''} - \varepsilon_k^2\delta_{k'',k'})\sum_{i,j=1}^{4} B_A^{(i)}(k'\sigma)n_F(\xi_{k'\sigma}^{(i)}) \cdot B_A^{(j)}(k''\sigma)n_F(\xi_{k''\sigma}^{(j)}) = t_1^2/(Ut_1 + 2\mu V_1)^2. \tag{13}$$

The critical pairing temperature is determined by the expression (13).

**3. Analyses and results**

The antiferromagnetic states can be found in these expressions. The Green's functions (8) give the electron number distributions with spin ↑ and ↓ in the forms

$$n_{A\uparrow}(k) = C_A^{(+)}(k\uparrow) n_F[E_A^{(+)}(k\uparrow)] - C_A^{(-)}(k\uparrow) n_F[E_A^{(-)}(k\uparrow)],$$

and

$$n_{A\downarrow}(k) = C_A^{(+)}(k\downarrow) n_F[E_A^{(+)}(k\downarrow)] - C_A^{(-)}(k\downarrow) n_F[E_A^{(-)}(k\downarrow)].$$

We get $S_A(k) = n_{A\uparrow}(k) - n_{A\downarrow}(k) > 0$, for $S_{l_A} > 0$ and the large $P_A$. The electrons of spin up in A-lattice have probability to occupy the states of spin down in the weak magnetic states. In the same way, take the translation $A \to B$, we get $n_{B\uparrow}(k) - n_{B\downarrow}(k) < 0$ for $S_{l_B} < 0$ and $S_{l_A} = -S_{l_B}$. The antiferromagnetic states are discovered in the Hubbard model. By the way, $S_{l_A} \neq 0$ require the spin-charge correlation $S_A \neq 0$, but $S_A \neq 0$ do not mean $S_{l_A} \neq 0$ (*for fully filled bands*). That is to say, the spin-charge correlation does not require appearing macroscopic magnetisms, in this case, the correlation means having short-range magnetic correlations. It is emphasized that the antiferromagnetism usually increases with the possible increasing of $S_A$.

The superconducting states is also shown when the spin-charge correlation exists in the model. In addition, the states below and around Fermi surface also contribute to the pairing.

If $P_A$ is very small, $S_A$ must be very small, all of the bands $E_{k\sigma}^{(i)}$ are overlap. The overlap states do not contribute to the superconductivity; this is because the overlap states do not contribute to the pairing functions on the basis of the equation (11). In another words, we find $B_A^{(i)}(k\sigma) n_F(\xi_{k\sigma}^{(i)}) = -B_A^{(j)}(k\sigma) n_F(\xi_{k\sigma}^{(j)})$, for $\xi_{k\sigma}^{(i)} = \xi_{k\sigma}^{(j)}$ ($i \neq j$), and we get $\sum_{i=1}^{4} B_A^{(i)}(k\sigma) n_F(\xi_{k\sigma}^{(i)}) = 0$ for the contributions of overlap states ($i, j = 1,2,3,4$.). That is to say, a very weak correlation can not allow superconductivity.

If $P_A$ is enough large but $S_A = 0$, superconductivity can not occur due to the overlaps between the bands $\xi_{k\sigma}^{(i)}$. If $S_A$ is enough large, the superconductivity may appear, it is easy to find there are the solutions of the temperature $T^{pair} > 0$ in (13). Particularly, because $\sum_{i,}^{4} B_A^{(i)}(k'\sigma) n_F(\xi_{k'\sigma}^{(i)})$ decreases with the increased temperature or the decreased $S_A$, the critical pairing temperature or the $T_c$ increase with the increased $S_A$. For example, this can be easily seen if $V_{ll'} = 0$.

It is shown that both superconductivity and antiferromagnetism can be included in the Hubbard model, therefore, the coexistence between the superconductivity and the weak antiferromagnetism is possible when the temperature is enough low. However, an exception is that the antiferromagnetism is increased with the decreasing temperature, the antiferromagnetism may exclude superconductivity.

## 4. Discussion

Our discussion can be extended to other problems. Firstly, the strong correlations have three kinds: the strong spin-spin correlation, the strong spin-charge correlation, and the strong charge-charge correlation, while the appropriate spin-charge correlation is a key role. As discussed above, both the antiferromagnetism and the superconductivity increase with the increased $S_A$. However, a weak antiferromagnetism is only considered in this work. In addition, $S_x$ and $S_y$ have not been considered in this work, hence it does not mean that the spin-charge correlation play the same role in magnetism and superconductivity. The strong spin-charge correlation can be divided into the short-range correlation and the long-range correlation. The former is dominated by the spin-spin correlation, while the latter is dominated by the charge-charge correlation. The short-range spin-charge correlation based superconductivity may be mediated by spin excitations, while the long-range spin-charge correlation based superconductivity may be mediated by charge excitations or phonons. We suggest that the underdoped cuprates are spin excitations and phonons mediated superconductors and the overdoped cuprates are charge excitations and phonons mediated superconductors for p-types of superconductors. These suggestions are consistent with their resistivity-temperature behaviors, and this has to be explained in other works. For example, spin correlation usually lead to $\rho \sim T^{-p}$ similar behavior ('localization of charges'), while charge correlation lead to the $\rho \sim T^{+p}$ behavior ('excitations of charges') in low temperatures. The spin excitations or the charge excitations should be from the nearly localized electrons. That is, there are two kinds of electron states in some materials; the nearly localized electrons provide the spin or charge excitations while the nearly free electrons provide the charge carriers. It seems to us that the appropriate spin-charge correlation is also necessary for the phonons mediated superconductivity. It is known that the materials such as Cu, Au, and Ag etc. do not show superconductivity. The BCS theory attributes the non-superconductivity to the extremely weak electron-phonon interaction, and hence the weak electron-phonon interaction leads to the little resistivity. We attribute the non-superconductivity to the extremely weak spin-charge correlation, and hence the weak correlation leads to the little resistivity. This is also understood because the weak spin-charge correlation means the weak scattering of charges by spins. The stronger spin-charge correlation is, the larger the resistivity is. In some aspects, our theory is similar to the BCS. For example, the BCS equations give correct but low $T_c$ on a single energy band of normal metals. Our theory also gives the low $T_c$ if our bands (9) form a band due to some overlap. In other aspects, our theory is different from the BCS. For example, the interactions $V_k$ may not be negative in Fermi surface, but the spin-charge correlation is required. The relations between the spin-charge correlation and the attractive force between electrons are next topics. Particularly, our theory can also explain the relation between magnetisms and superconductivity. For example, the features appearing the larger magnetic susceptibility $|\chi_m|$ are beneficial to superconductivity as shown in experiments, for which the spin-charge correlation may be stronger. In addition, if the magnetic field could strengthen the spin-charge correlation, the magnetic field induced superconductivity [9] may occur, while this behavior could not be explained with BCS.

Secondly, the spectral weight transfer [13] could be explained in this work. If $P_A$ is enough large, but $S_A$=0, the state density around the Fermi surface is zero for the half-filled model. We can understand that the spin-charge correlation is increased with increased doping in cuprates, hence the states around the Fermi surface appear with doping as shown in (8). The "increased spin-charge correlation" is related to the increased charge-charge correlation. Thirdly, the gap-function is related to the $V_{A\Delta}^+(k\sigma)$, the s- and d-wave symmetry [14] can be found in (10) for an anisotropic model. However, $V_{A\Delta}^+(k\sigma)$ is the anisotropic s-wave symmetry if the nearest-neighbor approximation is taken as $t_{ll'} = t$. Fourthly, the unusual isotope effects (on $T_c$) in optimally doped cuprates can be understood [15], because the strong correlation may dominate the superconductivity in this region. As discussed above, a higher $T_c$ is due to the higher $S_A$. Fifthly, the *T*-linear resistivity of cuprate superconductors may be due to the charge excitations [16], this is because the higher $S_A$ may be dominated by the charge-charge correlation, and the charge correlation can lead to the charge excitations. In addition, if weak antiferromagnetism exists, with the expression (13) we find $F_{AA}^+(k\uparrow) > F_{AA}^+(k\downarrow)$ and $F_{BB}^+(k\downarrow) > F_{BB}^+(k\uparrow)$ in the amounts of these functions, this is because these electrons in A-lattice have a larger probability to be spin up, while these electrons in B-lattice have a larger probability to be spin down.

In summary, the superconducting critical temperature increases with the strengthened spin-charge correlation, and many properties of materials are related to the spin-charge correlation.